# Direct measurements of the energy flux due to chemical reactions at the surface of a silicon sample interacting with a $SF_6$ plasma

R. Dussart[a)], A. L. Thomann, L. E. Pichon, L. Bedra, N. Semmar, P. Lefaucheux, J. Mathias, Y. Tessier

*GREMI-Polytech'Orléans, 14 rue d'Issoudun, BP 6744, 45067 Orléans cedex 2, France*

## Abstract

Energy exchanges due to chemical reactions between a silicon surface and a $SF_6$ plasma were directly measured using a heat flux microsensor. The energy flux evolution was compared with those obtained when only few reactions occur at the surface to show the part of chemical reactions. At 800 W, the measured energy flux due to chemical reactions is estimated at about 7 W.cm$^{-2}$ against 0.4 W.cm$^{-2}$ for ion bombardment and other contributions. Time evolution of the HFM signal is also studied. The molar enthalpy of the reaction giving $SiF_4$ molecules was evaluated and is consistent with values given in literature.

[a)] E-mail : Remi.Dussart@univ-orleans.fr

The energy transfer from a plasma to a surface is usually evaluated by indirect measurements (thermocouples, evaluation from calculations…).[1-3] Direct measurements have been reported using Gardon gauge for deposition[4] and etching[5] experiments. Another technique based on scanning calorimetry is also described in references [6-7] to determine the heat due to plasmochemical reactions.

Three different types of species have to be taken into account to evaluate the energy involved during the interaction with the surface: charge carriers, neutrals and photons.[1] Direct evaluation of the energy flux density is a real improvement especially in plasma processes like etching and deposition.[8,9] We know that the energy transfer is quite significant in some particular plasma processes like cryoetching for example where chemical reactions can play an important role in terms of energy flux. We have already reported on the effect of the energy released by exothermic chemical reactions, which are involved in the silicon etching by fluorine atoms. This deposited energy seems to be a non negligible cause of the extraction of the passivation layer.[10,11]

A Heat Flux Microsensor (HFM) (Vattel HFM7-Vattel@) was installed in order to directly measure the global energy transfer of the plasma to a surface[12]. This microsensor is based on the Seebeck effect. It is composed of a Pt 100 temperature sensor and hundreds of micro thermocouples (1600 $cm^{-2}$) at the surface covered by an absorbent material (Zynolite).[13] The HFM was calibrated using a NIST protocol, which was carried out with a home made black body[12,14]. The active part of the HFM has a diameter of 6 mm.

Results obtained in Ar plasma without sample in an ICP reactor at low pressure (1 to 10 Pa) were recently presented.[12,15] A plasma study was carried out to estimate the contribution of the charge carriers in the global energy transfer[12]. Gas temperature and metastable density were measured by tuneable diode laser absorption experiments to evaluate the contribution of the gas conduction[15]. The potential energy, which metastable atoms could release at the surface by collisions, was also estimated. The result of this study showed that, in our experimental conditions, the main contribution was due to ions even if the surface was not biased. The gas conduction was not negligible (about 10 mW.cm$^{-2}$ for a 500 W Ar plasma at 1 Pa), although it represented about a tenth of the global energy transfer in our experimental conditions[15]. The energy flux density released by metastable recombination at the surface was found negligible especially in inductive mode for which the metastable density is reduced by quenching with electrons[15].

In this letter, we present results obtained when a plasma of $SF_6$ interacts with a silicon sample. This experiment was made to directly estimate the energy flux due to chemical reactions of fluorine with silicon. An experiment based on an indirect estimation of the temperature surface had been carried out by another team few years ago to show the increase of the silicon surface temperature during etching[16]. In $SF_6$ plasma, the main by-product of silicon etching at room temperature is $SiF_4$[17,18]. The chemical reaction to form this by-product is strongly exothermic[18].

A sample of about 1 cm² (100) silicon was directly mounted on the HFM. A thermal paste (HTCA 200 from Electrolube)[19] was used to both maintain the sample in contact with the HFM (without the need of screws) and ensure a good thermal contact between them. This paste has a rather good thermal conductivity of 0.9 W.m$^{-1}$.K$^{-1}$. Measurements were performed at about 30 cm below the inductively coupled plasma source in the middle of the diffusion chamber. The HFM was cooled down to 5°C by water flow and remained at this temperature during the plasma treatment. The HFM voltage was recorded with a nano-voltmeter (Keihtley 2182) every 0.5 second.

In figure 1, we present the results obtained in a pure Ar plasma (20 sccm, 3 Pa) for different source powers without any sample. Figure 1.a represents the HFM signal versus time. When the plasma is switched on, a significant increase of the HFM signal is observed within one second. Then the signal slowly increases due to the heating of the reactor walls (radiative contribution). Figure 1b gives the energy flux density versus plasma source power. A value of 350 mW.cm$^{-2}$ is obtained for 1200 W source power. This energy flux density is mostly due to the ion bombardment at the surface of the HFM[12]. Langmuir probe measurements were carried out in the same conditions of plasma and give the values of the ion density ($n_i = 1,1 \times 10^{11}$ cm$^{-3}$), the electron temperature ($T_e = 2,3$ eV), the plasma potential ($V_p = 25$ V) and the floating potential ($V_f = 15$ V).

Figure 2a shows the energy flux density directly measured when a plasma of $SF_6$ (800 W – 3 Pa – 20 sccm) is in interaction either with a bulk silicon sample or with an oxidized silicon sample set to the surface of the HFM. We also added (figures 2a and 2b) the signal obtained if the oxidized sample is in interaction with an Ar plasma (800 W – 3 Pa – 20 sccm).

In the case of the bulk silicon interacting with the $SF_6$ plasma, we first observed a sharp increase (within 2 sec) of the HFM signal immediately after the plasma has been switched on. Then we observe a rather exponential increase of the signal, which corresponds to the heating of the sample. We attribute the sharp and significant increase of the HFM signal to the energy flux released by the chemical reactions between fluorine and silicon to form the $SiF_4$ volatile molecules. This is checked by the absence of this sharp increase of the signal when no or only few reactions occur at the surface of the sample, which is the case when a not biased $SiO_2$ surface interacts with an $SF_6$ plasma.[20] In this case, the increase is exponential and the measured energy flux density is quite close to the one obtained when the substrate is submitted to a non reactive argon plasma. Again this exponential increase is attributed to the heat of the sample as it will be described further.

In figure 3, the results obtained in the case of a silicon sample submitted to an $SF_6$ or an Ar plasma at different source powers are presented. Figure 3a gives an example of the HFM signal for different plasma source powers. The maximum energy flux density is shown in figure 3b versus plasma power for $SF_6$ plasma interacting with Si sample, and compared to the cases where no reaction can occur ($SF_6$ plasma on $SiO_2$, Ar plasma on $SiO_2$ and Ar plasma on Silicon). In the case of $SF_6$ plasma on silicon sample, we can observe a change of the curve evolution between 300 and 500 W. It corresponds to the transition between capacitive and inductive mode, for which the $SF_6$ dissociation is significantly enhanced producing much more fluorine atoms. In the three other cases, the maximum energy flux density is much smaller and corresponds to the heating of the surface mainly due to ion bombardment.

The time evolution of signals shown in figure 2b (no or very few reactions between the plasma and the surface) can be explained as follows. If we assume that the temperature T is uniform within the sample at any instant during the transient process and if we consider that the main energy flux comes from the plasma and is evacuated by the HFM, the energy balance gives:

$$F_0 S_1 dt - \left(\frac{T - T_{HFM}}{R}\right) dt = m C_p dT \qquad (1)$$

$F_0$ is the energy flux density coming from the plasma

$T_{HFM}$ is the temperature of the HFM, which is monitored at 5°C.

R is the contact resistance between the HFM and the silicon sample.

T is the silicon sample temperature

m is the mass of the sample

$C_p$ is the calorific capacity of silicon (712 $J.kg^{-1}.K^{-1}$).

$S_1$ is the surface in interaction with the plasma (silicon surface and mechanical clamping).

$S_2$ is the active surface of the HFM.

By resolving equation (1), the temperature time distribution is easily established:

$$T = R F_0 S_1 (1 - e^{-\frac{t}{\tau}}) + T_{HFM} \quad \text{with} \quad \tau = R m C_p \qquad (2)$$

which finally gives the time evolution of the energy flux density:

$$F = \frac{1}{RS_2}(T - T_{HFM}) = \frac{S_1}{S_2} F_0 (1 - e^{-\frac{t}{\tau}}) \qquad (3)$$

This simple model can give us an order of magnitude of the contact resistance between the sample and the HFM. When we fit the curves of figure 2 using the relation (3), we find a characteristic time $\tau$ of $7 \pm 3$ seconds. This value gives a contact resistance of $75 \pm 25$ K.W$^{-1}$, which is assumed constant during the transient regime. This is consistent with a paste thickness of the order of 1 mm, which typically corresponds to the quantity we spread on the sample backside.

It is possible to infer the enthalpy for the reaction from these direct measurements. An averaged silicon etch rate as high as $3.8 \pm 0.2$ µm.min$^{-1}$ was determined by SEM measurements of the sample after 80 min of SF$_6$ plasma (800 W, 3 Pa, 20 sccm). Since the silicon surface exposed to the plasma is rather large ($1.7 \pm 0.1$ cm$^2$), we do not have any Aspect Ratio Dependent Etching effect[21], and we can consider that the etch rate remains constant during the plasma process. The expression of the enthalpy for this reaction can be estimated by the following relation (4):

$$H_r = \frac{\Phi M_{S_i} S_2}{\rho_{S_i} v_g S_1} \qquad (4)$$

where $H_r$ is the molar enthalpy for the reaction in J.mol$^{-1}$.

$\Phi$ is the energy flux density due to reactions directly measured by the HFM (in W.m$^{-2}$).

$M_{Si}$ is the molar mass of silicon (28.0855 g.mol$^{-1}$)

$\rho_{S_i}$ is the volumic mass of silicon (2.329x10$^6$ g.m$^{-3}$).

$v_g$ is the etch rate (m.s$^{-1}$)

$S_1$ is the total silicon surface exposed to the plasma

$S_2$ is the active surface of the sensor

If we consider that the energy flux due to chemical reaction is constant, we can estimate the enthalpy for the reaction. The energy flux density due to chemical reactions is estimated at about 7000 ± 500 mW.cm$^{-2}$ during the SF$_6$ plasma at 800 W, 3 Pa, 20 sccm. This estimation is made after removing the exponential growth due to the plasma heating. With this value, we calculate a molar enthalpy of about -2200 ± 400 kJ.mol$^{-1}$ for an etch rate of 3.8 µm.min$^{-1}$. The value of the standard enthalpy for the chemical reaction of silicon etching by fluorine atoms (Si + 4F $\rightarrow$ SiF$_4$) given in literature[12] (25°C, 1 bar) is -1931 kJ.mol$^{-1}$, which is in agreement with our estimation.

The temperature increase can modify a little bit the tabulated value of the molar enthalpy given at 25°C. Although, SiF$_4$ is known as the major by product in silicon etching, it is also known that SiF$_2$ production can be significant[22] especially at higher temperature[18]. The molar enthalpy to form SiF$_2$ is quite different from the one corresponding to SiF$_4$ formation[18]. This comparison between the value inferred from our direct measurements and the tabulated value, allows us to check that we obtain the same order of magnitude even if an estimation taking into account the SiF$_2$ proportion, temperature dependence… would give a better accuracy to this comparison.

This heat flux microsensor is much more convenient than a simple temperature sensor, because it provides a direct measurement and it is able to instantaneously give the heat flux due to either the plasma and/or the chemical reactions occurring at the surface. The results are in very good agreement with tabulated values. The sample temperature measurement alone would not have given the energy flux density, which makes this diagnostic very powerful for plasma processes, especially in etching (end point detection …) or for other types of processes for which chemical reactions occur at the surface (surface modification, deposition…)

**Figure Captions**

Figure 1 : (a) Signal of the HFM without sample measured in an Ar plasma for different source powers versus time

        (b) Energy flux density as a function of the plasma source power

Figure 2: Energy flux measured during the plasma interaction with a bulk silicon sample or with a silicon sample covered by a thermal $SiO_2$ layer (1 µm thick)

        (a) Comparisons between the signal obtained when a silicon sample is submitted to the $SF_6$ plasma (800 W, 20 sccm, 3 Pa) and the signal obtained when an oxidized silicon sample is submitted to a $SF_6$ plasma (800 W, 20 sccm, 3 Pa) or an Ar plasma (800 W, 20 sccm, 3 Pa)

        (b) Zoom of the results shown in figure 2a (lower part of the graph) for which an $SF_6$ plasma or an Ar plasma interacts with a $SiO_2$ surface.

Figure 3: (a) Time evolution of the HFM signal during an Ar plasma followed by an $SF_6$ plasma in interaction with a silicon sample for different source powers versus time. Experimental conditions: 20 sccm of $SF_6$ (or Ar) – 3 Pa

        (b) Maximum energy flux density versus the source power density obtained for samples of Silicon or oxidized silicon in interaction with $SF_6$ plasma or with Ar plasma.

Fig. 1

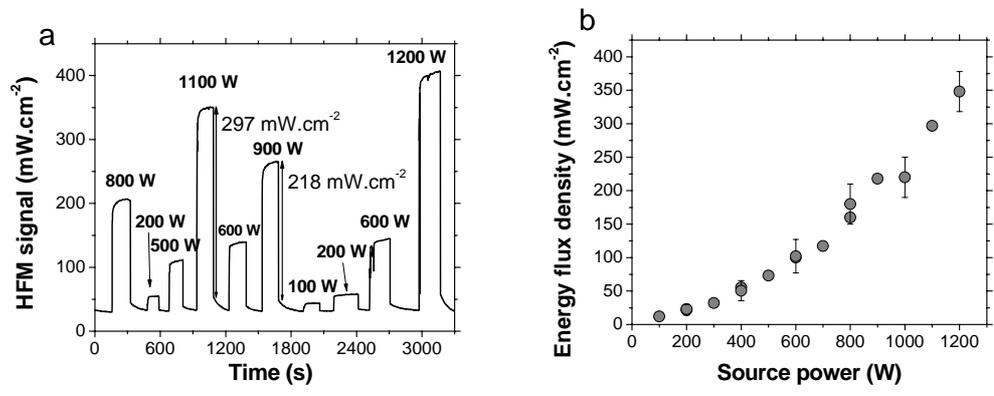

Fig. 2

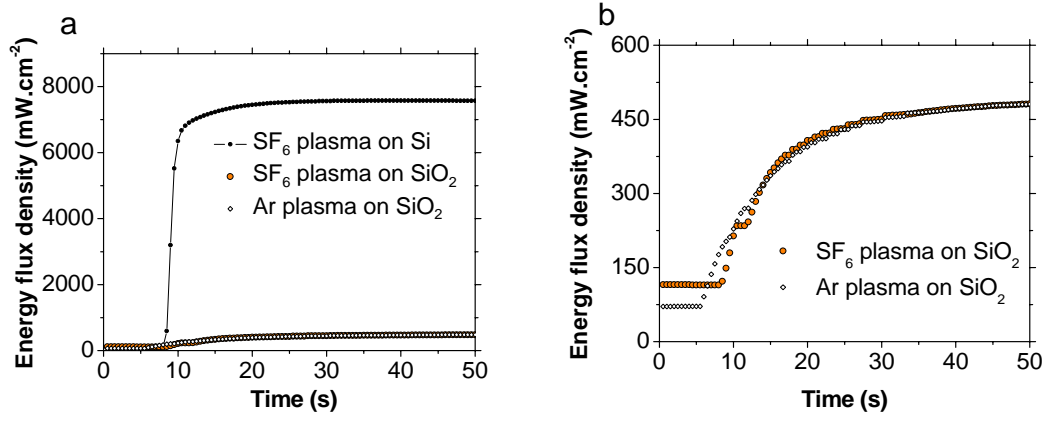

Fig. 3

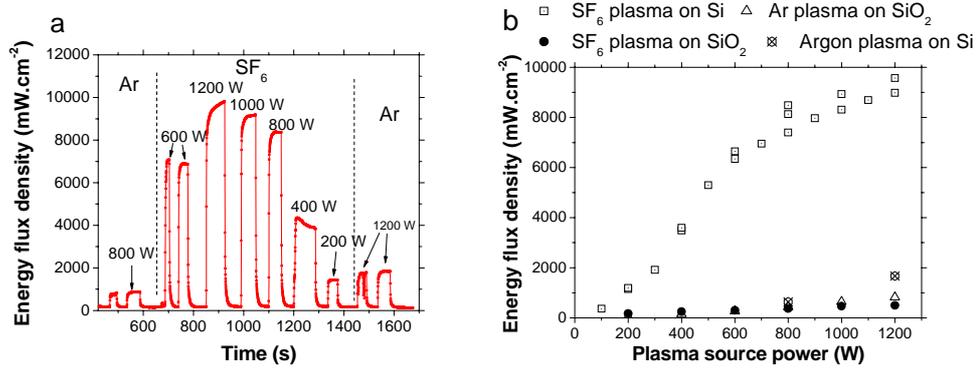